\documentclass{egs}
\usepackage{times,textcomp}
\usepackage{graphicx}

\printfigures

\begin{document}
\title{Cross Recurrence Plot Based Synchronization of Time Series}
\author[1]{N. Marwan}
\author[1]{M. Thiel}
\affil[1]{Institute of Physics, University of 
        Potsdam, Germany}
\author[2]{N.\,R. Nowaczyk}
\affil[2]{GeoForschungsZentrum Potsdam, Germany}

\date{Manuscript final version from November 29, 2001}
\journal{\NPG}
\firstauthor{Marwan}
\proofs{N.\,Marwan\\Institute of Physics\\University of Potsdam\\14415 Potsdam\\Germany}
\offsets{N.\,Marwan\\Institute of Physics\\University of Potsdam\\14415 Potsdam\\Germany}

\msnumber{1040}

\maketitle

\begin{abstract}
The method of recurrence plots is extended to the cross
recurrence plots (CRP), which among others enables the study 
of synchronization
or time differences in two time series. This is emphasized
in a distorted main diagonal in the cross recurrence plot, the
line of synchronization (LOS). A non-parametrical fit of this LOS
can be used to rescale the time axis of the two data series 
(whereby one of it is e.\,g.~compressed or stretched) so that 
they are synchronized. An application of this method to
geophysical sediment core data illustrates its suitability for
real data. The rock magnetic data of two different sediment 
cores from the Makarov Basin can be adjusted to each other
by using this method, so that they are comparable.
 \end{abstract}

\section{Introduction}
The adjustment of data sets with various time scales occurs in
many occasions, e.\,g.~data preparation of
tree rings or geophysical profiles.
In geology, often a large set of geophysical data series is gained at
various locations (e.\,g.~sediment cores).
That is why these data series have a different length and time scale.
Before any time series analysis can be started, the data series
have to be synchronized to the same time scale.
Usually, this is done visually by comparing and
correlating each maximum and minimum in both data sets by hand
(wiggle matching), which includes the human factor of 
subjectiveness and is a lengthy process. An automatical and 
objective method for verification should be very welcome. 

In the last decades some techniques for this kind of
correlation and adjustment were suggested. They span
graphical methods \citep{prell86}, inverse algorithms, 
e.\,g.~using Fourier series \citep{martinson82} and 
algorithms based on similarity of data, e.\,g.~sequence slotting
\citep{thompson89}.

However, we focus on a method based on nonlinear time series 
analysis. During our investigations of the method of
cross recurrence plots (CRP), we have found an interesting
feature of it. Besides the possibility of the application of 
the recurrence quantification analysis (RQA) of Webber and 
Zbilut on CRPs (\citeyear{zbilut98}), 
there is a more fundamental
relation between the structures in the CRP and the 
considered systems. Finally, this feature can be used
for the task of the synchronization of data sets.
Although the first steps of this method are similar to 
the sequence slotting method, their roots are different. 

First we give an introduction in CRPs. Then we
explain the relationship between the structures in the CRP
and the systems and illustrate this with a simple model.
Finally we apply the CRP on geophysical data in order
to synchronize various profiles and to show their
practical availability. Since we focus on the synchronization
feature of the CRP, we will not give a comparison between
the different alignment methods.

\section{The Recurrence Plot}

Recurrence plots (RP) were firstly introduced by \citet{eckmann87}
in order to visualize time dependent behaviour
of orbits $\vec x_i$ in the phase space. 
A RP represents the recurrence of the phase space trajectory to a
state. The recurrence of states is a fundamental property
of deterministic dynamical systems \citep{argyris94, casdagli97,kantz97}. 
The main step of the visualization is the calculation of the
$N \times N$-matrix   

\begin{equation}\label{eqRP}
\mathbf{R}_{i,\,j} = \Theta\bigl(\varepsilon-\|\vec x_{i} - \vec x_{j}\|\bigr), \quad \, 
i, j=1\dots N,
\end{equation}

where $\varepsilon$ is a predefined 
cut-off distance, $\| \cdot \|$ is the norm (e.\,g.~the Euclidean norm) and
$\Theta(x)$ is the Heaviside function. The values
{\it one} and {\it zero} in this matrix can be simply 
visualized by the colours black and white. 
Depending on the kind of the application, $\varepsilon$ can be a fixed
value or it can be changed for each $i$
in such a way that in the ball with the radius $\varepsilon$
a predefined amount of neighbours occurs. The latter will provide a constant 
density of recurrence points in each column of the RP.

The recurrence plot exhibits characteristic patterns
for typical dynamical behaviour \citep{eckmann87,webber94}: 
A collection of single recurrence points, homogeneously and irregularly 
distributed over the whole plot, reveals a mainly stochastic process.
Longer, parallel diagonals formed by recurrence points and with the same 
distance between the diagonals are caused by periodic processes. 
A paling of the RP away from the main diagonal to the corners reveals
a drift in the amplitude of the system. 
Vertical and horizontal white bands in the
RP result from states which occur rarely or represent extreme states.
Extended horizontal and vertical black lines or areas occur if a state
does not change for some time, e.\,g.~laminar states. 
All these structures were formed
by using the property of recurrence of states. 
It should be pointed out that the states are
only the ``same'' and recur in the sense of the vicinity, which is
determined by the distance $\varepsilon$. RPs and their quantitative
analysis (RQA) became more well known in the last decade \citep[e.\,g.][]{casdagli97}. 
Their applications
to a wide field of miscellaneous research show their suitability in
the analysis of short and non-stationary data.

\section{The Cross Recurrence Plot}

Analogous to \citet{zbilut98}, we have expanded the
method of recurrence plots (RP) to the method of {\it cross recurrence plots}. 
In contrast to the conventional RP, two time
series are simultaneously embedded in the same phase space. The test
for closeness of each point of the first trajectory $x_i$ ($i=1 \dots N$) 
with each point of the second trajectory $y_j$ ($j=1 \dots M$)
results in a $N \times M$ array

\begin{equation}\label{eqCRP}
\mathbf{CR}_{i,\,j}=\Theta\bigl(\varepsilon-\|\vec x_i-\vec y_j\|\bigr).
\end{equation}

Its visualization is called the {\it cross recurrence plot (CRP)}. 
The definition of the closeness between both trajectories can
be varied as described above. Varying $\varepsilon$ may be 
useful to handle systems with different amplitudes.

The CRP compares the considered systems and allows us to benchmark the
similarity of states.
In this paper we focus on the bowed ``{\it main diagonal}'' 
in the CRP, because it is related to the frequencies and
phases of the systems considered.

\section{The Line of Synchronization in the CRP}

Regarding the conventional RP, Eq.~\ref{eqRP},
one always finds a main diagonal in the plot, because of the identity 
of the $(i,i)$-states. The RP can be considered as a special case of 
the CRP, Eq.~\ref{eqCRP}, which usually does not have a main diagonal as the 
$(i,i)$-states are not identical. 

In data analysis one is often faced with time series that are 
measured on varying time scales.
These could be sets from borehole or core data in geophysics or tree
rings in dendrochronology. Sediment cores might have undergone a number of
coring disturbances such as compression or stretching. Moreover, cores from
different sites with differing sedimentation rates would have different
temporal resolutions. All these factors require a method of synchronizing.

\begin{figure}
  \includegraphics[width=7.5cm]{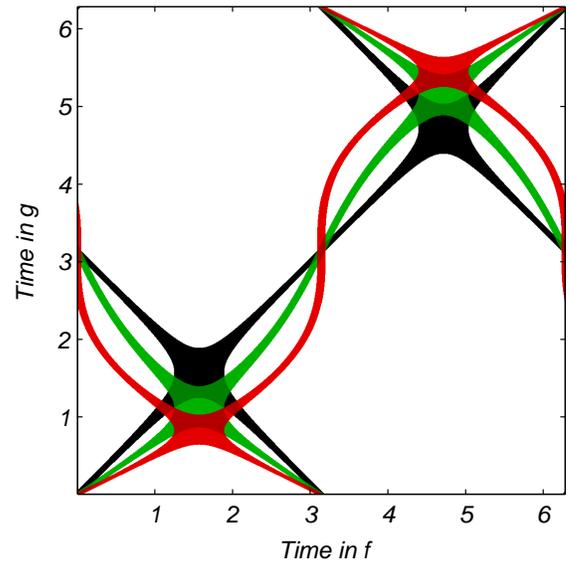}
  \caption[]{\label{crp.variation}
Cross recurrence plots of sine functions $f(t)=\sin(\varphi t)$
and $g(t)=\sin(\varphi t + a  \sin(\psi t))$, 
whereat $a=0$ for the black CRP, $a=0.5$ for the green CRP and
$a=1$ for the red CRP. The variation in the time domain
leads to a deforming of the synchronization line.}
\end{figure}

A CRP of the two corresponding 
time series will not contain a main diagonal. But if the sets of 
data are similar e.\,g.~only rescaled, a more or less continuous 
line in the CRP that is like a distorted main diagonal can occur. 
This line contains information on the rescaling. 
We give an illustrative example. A CRP of a sine function 
with itself (i.\,e.~this is the RP) contains a main diagonal
(black CRP in Fig.\,\ref{crp.variation}). Hence, the CRPs in 
the Fig.\,\ref{crp.variation} are computed with embeddings of
dimension one, further diagonal lines from the upper left to
the lower right occur. These lines typify the similarity of
the phase space trajectories in positive and negative time direction.

Now we rescale the time axis of the second sine function in the following way
\begin{equation}
\sin (\varphi t) \longrightarrow \, \sin \bigl(\varphi t + a  \sin(\psi t)\bigr)
\end{equation}

We will henceforth use the notion rescaling only in the mention
of the rescaling of the time scale.
The rescaling of the second sine function with different parameters $\varphi$ 
results in a deformation of the main diagonal 
(green and red CRP in Fig.\,\ref{crp.variation}). The distorted line 
contains the information on the rescaling, which we will need in order to 
re-synchronize the two time series. Therefore, we call this 
distorted diagonal {\it line of synchronization (LOS)}.
  
In the following, we present a toy function to explain the procedure. 
If we consider a one dimensional case without embedding, the CRP is
computed with
\begin{equation}\label{Bed1}
\mathbf{CR}(t_1, t_2)=\Theta\bigl(\varepsilon-\left\| f(t_1)-g(t_2)\right\|\big) .
\end{equation}

If we set $\varepsilon=0$ to simplify the condition, 
Eq.\,(\ref{Bed1}) delivers a recurrence point if 
\begin{equation}\label{Bed2}
f(t_1)=g(t_2).
\end{equation}
In general, this is an implicit condition that links the variable 
$t_1$ to $t_2$.
Considering the physical examples of above, it can be assumed that the time 
series are essentially the same -- that means that $f=g$ -- up to a rescaling 
function of time. 
So we can state
\begin{equation}\label{Bed3}
f(t_1)=f\bigl(\phi(t_1)\bigr).
\end{equation}
If the functions $f(\cdot)$ and $g(\cdot)$ are not identical our method is in 
general not capable of deciding if the difference in the time series is due to 
different dynamics ($f(\cdot)\ne g(\cdot)$) or if it is due to simple 
rescaling. So the assumption that the dynamics is alike up to a rescaling 
in time is essential. Even though for some cases where $f \ne g$ it can be 
applied in the same way. If we consider the functions 
$f(\cdot)=a\cdot \bar{f}(\cdot) +b$ and $g(\cdot)=\bar{g}(\cdot)$,
whereby $f(\cdot) \ne g(\cdot)$  are the observations and 
$\bar{f}(\cdot) = \bar{g}(\cdot)$ are the states,
normalization with respect to the mean and the standard deviation allows 
us to use our method.
\begin{eqnarray}
f(\cdot) &=& a\cdot \bar{f}(\cdot)+b \longrightarrow \tilde{f}(\cdot)=\frac{f(\cdot)-\langle f(\cdot)\rangle}{\sigma\left(f(\cdot)\right)}\\
\tilde{g}(\cdot) &=& \frac{g(\cdot)-\langle g(\cdot)\rangle}{\sigma\left(g(\cdot)\right)}
\end{eqnarray}
With $\bar{g}(\cdot)=\bar{f}(\cdot)$ the functions $\tilde{f}(\cdot)$ and 
$\tilde{g}(\cdot)$ are the same after the normalization. Then our method 
can be applied without any further modification.

In some special cases Eq.\,(\ref{Bed3}) can be resolved with respect 
to $t_1$. Such a case is a system of two sine functions with different 
frequencies
\begin{eqnarray}
f(t) &=& \sin(\varphi\cdot t+\alpha)\\
g(t) &=& \sin(\psi\cdot t+\beta)
\end{eqnarray}
Using  Eq.\,(\ref{Bed2}) and Eq.\,(\ref{Bed3}) we find
\begin{equation}
\sin \left( \varphi\, t_1 + \alpha\right) - \sin \left( \psi\,t_2 + \beta \right) = 0 \\
\end{equation}
and one explicit solution of this equation is
\begin{equation}
\Rightarrow \quad t_2= \phi(t_1)= \left( \frac{ \varphi}{\psi} t_1 + \gamma\right) 
\end{equation}
with $\gamma = \frac{\alpha-\beta}{\psi}$ .
In this special case the slope of the main line in a cross recurrence plot 
represents the frequency ratio and the distance between the axes origin 
and the intersection of the line of synchronization with the ordinate 
reveals the phase difference.
The function $t_2=\phi(t_1)$ (Eq.\,\ref{Bed3}) is a transfer or rescaling 
function which allows us to rescale the second system to the first system. 
If the rescaling function is not linear the LOS will also be curved.

For the application one has to determine the LOS -- usually non-parametrically --
and then rescale one of the time series. In the Appendix we describe
a simple algorithm for estimating the LOS. Its determination will
be better for higher embeddings because the vertical and
cross-diagonal structures will vanish. We do not conceal that
the embedding of the time series is connected with difficulties.
The Takens Embedding Theorem holds for closed, deterministic systems without noise, 
only. If noise is present one needs its realization to find a reasonable embedding. 
For stochastic time series it does not make sense to consider a phase space and so 
embedding is in general not justified here either \citep{romano01,takens81}. 

The choice of a special embedding lag could be correct for one section of the data,
but incorrect for another 
(for an example see below). This can be the case if the data is non-stationary. 
Furthermore, the choice of
method of computing the CRP and the threshold $\varepsilon$ will
influence the quality of the estimated LOS.

The next sections will be dedicated to application.

\section{Application to a Simple Example}

\begin{figure}
  \includegraphics[width=7.5cm]{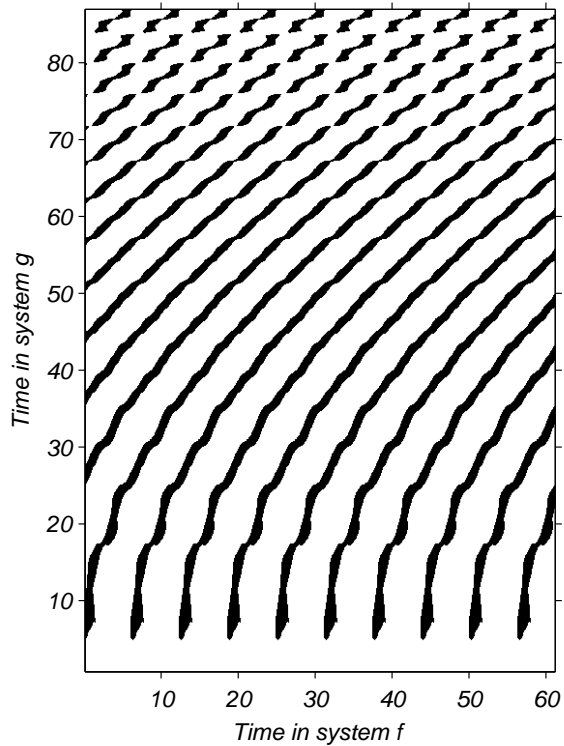}
  \caption[]{\label{crp.bsp1}
Cross recurrence plots of two sine functions $f(t)=\sin(\varphi t)$
and $g(t)=\sin(\psi t^2))$ which is the base for the determination
of the rescaling function between both data series. The embedding parameters
were dimension $m=2$, delay $\tau=\pi/2$ and a varying threshold $\varepsilon$,
in such a way that the CRP contains a constant recurrence density of 20\,\%.}
\end{figure}

\begin{figure}
  \includegraphics[width=8.2cm]{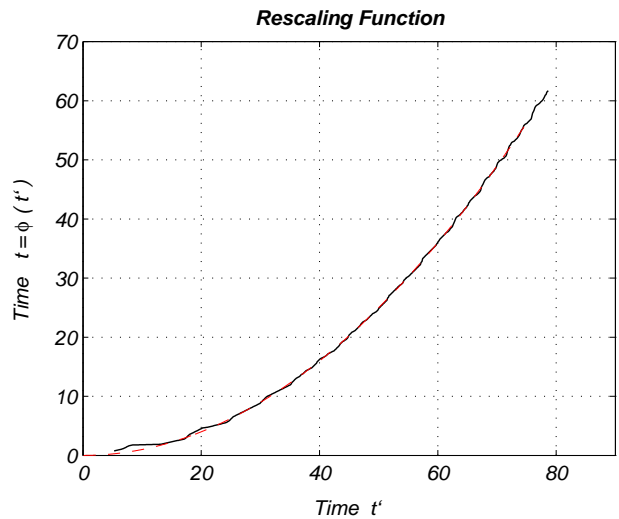}
  \caption[]{\label{trans.bsp1}
The rescaling function (black) determined from the CRP in 
Fig.~\ref{crp.bsp1}. It has the expected parabolic shape of the
squared coherence in the time domain. In red the square 
function.}
\end{figure}

\begin{figure}
  \includegraphics[width=8.3cm]{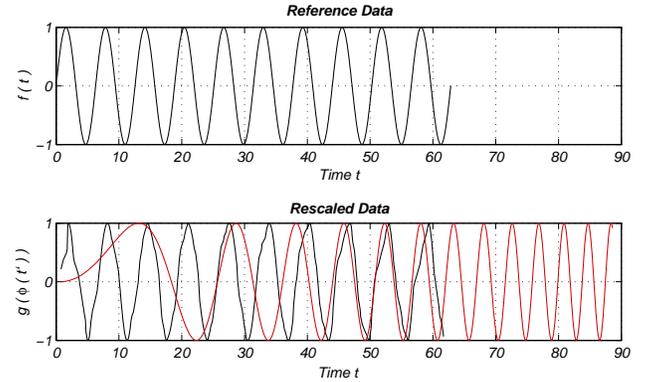}
  \caption[]{\label{rescaled.bsp1}
Reference data series (upper panel) and rescaled data 
series before (red) and after (black) rescaling by using 
the rescaling function of Fig.~\ref{trans.bsp1} (lower panel).}
\end{figure}

At first, we consider two sine functions $f(t)=\sin(\varphi t)$
and $g(t)=\sin(\psi t^2)$, where the time scale
of the second sine differs from the first by a quadratic term and
the frequency $\psi=0.01 \, \varphi$.
Sediment parameters are related to such kind of functions
because gravity pressure increases nonlinearly with the depth.
It can be assumed that both data series come from the same process and were
subjected to different deposital compressions (e.\,g.~a squared
or exponential
increasing of the compression). Their CRP contains a bowed
LOS (Fig.\,\ref{crp.bsp1}). We have used the embedding parameters
dimension $m=2$, delay $\tau=\pi/2$ and a varying threshold $\varepsilon$,
so that the CRP contains a constant recurrence density of 20\,\%. 
Assuming that the time scale of $g$ is not the correct scale, we denote
that scale by $t'$. In order to determine the
non-parametrical LOS, we have implemented the algorithm described
in the Appendix. Although this algorithm is still not mature, we
obtained reliable results (Fig.\,\ref{trans.bsp1}). The resulting rescaling 
function has the expected squared shape $t=\phi(t')=0.01 \, t'^2$ (red curve in 
Fig.\,\ref{trans.bsp1}). Substituting the time scale $t'$
in the second data series $g(t')$ by this rescaling function 
$t=\phi(t')$, we get a set of synchronized data $f(t)$ and $g(t)$
with the non-parametric rescaling function $t=\phi(t')$
(Fig.\,\ref{rescaled.bsp1}). The synchronized data series are
approximately the same. The cause of some differences is the 
meandering of the LOS, which itself is caused by partial weak embedding.
Nevertheless, this can be avoided by more complex algorithm for
estimating the LOS.

\section{Application to Real Data}

In order to continue the illustration of the working of our method
we have applied it to real data from geology.

\begin{figure}
  \includegraphics[width=8.3cm]{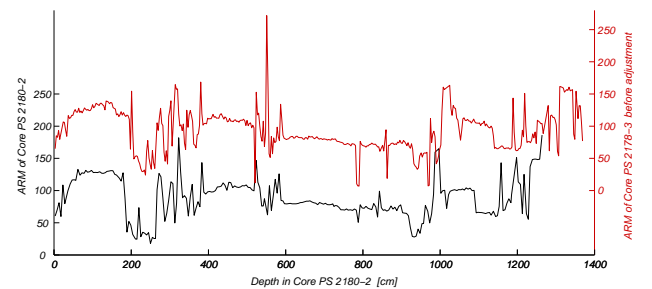}
  \caption[]{\label{arm_ba}
ARM data of the boreholes PS~2178-3~GPC and PS~2180-2~GPC
in the Central Arctic Ocean before adjustment.}
\end{figure}

In the following we compare the method of cross recurrence plot matching
with the conventional method of visual wiggle matching (interactive adjustment).
Geophysical data of
two sediment cores from the Makarov Basin, central Arctic Ocean, PS~2178-3
and PS~2180-2, were analysed. The task should be to adjust
the data of the PS~2178-3 data (data length $N=436$) 
to the scale of the PS~2180-2 (data length $N=251$)
in order to get a depth-depth-function which allows to
synchronize both data sets (Fig.\,\ref{arm_ba}).

We have constructed the phase space with the normalized 
six parameters low field magnetic susceptibility ($\kappa_{L\!F}$),
anhysteretic remanent magnetization ($ARM$), 
ratio of anhysteretic susceptibility to $\kappa_{L\!F}$
($\kappa_{ARM}/\kappa_{L\!F}$), relative palaeointensity ($P\!J\!A$),
median destructive field of $ARM$ ($M\!D\!F_{ARM}$) and
inclination ($I\!N\!C$). A comprehensive discussion of 
the data is given in \cite{nowa2001}. The embedding was combined 
with the time-delayed method according to \citep{takens81} in order to
increase further the dimension of the phase-space  
with the following rule: If we have $n$ parameters $a_i$, 
the embedding with dimension $m$ and delay $\tau$ will result
in a ($m\cdot n$)-dimensional phase space:
\begin{eqnarray}
\vec x(t) &=& \big( a_1(t), \dots, a_n(t), \nonumber \\
          & &{} a_1(t+\tau), \dots, a_n(t+\tau),\nonumber \\
          & &{} a_1(t+2\tau), \dots, a_n(t+2\tau), \dots \nonumber \\
	  & &{}	a_1(t+(m-1)\tau), \dots, a_n(t+(m-1)\tau \big)
\end{eqnarray}

For our investigation we have used a dimension $m=3$ and a 
delay $\tau=1$, which finally led
to a phase space of dimension 18 ($3\times 6$). The recurrence
criterion was $\varepsilon=5$\,\% nearest neighbours.

\begin{figure}
  \includegraphics[width=8.3cm]{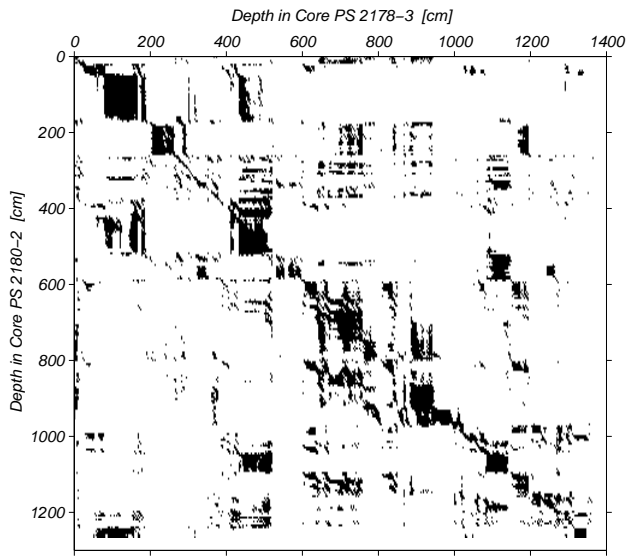}
  \caption[]{\label{crp}
Cross recurrence plot based on six normalized sediment parameters 
and an additional embedding dimension of $m=3$ ($\tau=1$,
$\varepsilon=0.05$).}
\end{figure}

\begin{figure}
  \includegraphics[width=8.3cm]{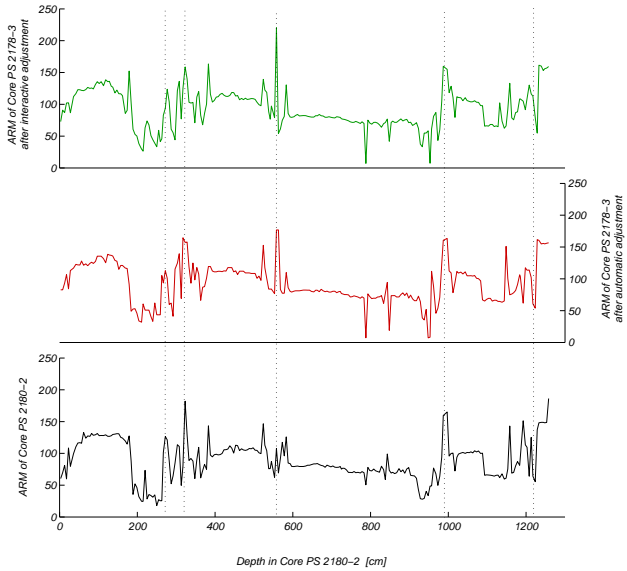}
  \caption[]{\label{arm_aa}
ARM data after adjustment by wiggle matching (top)
and by automatic adjustment using the LOS from Fig.~\ref{crp}.
The bottom figure shows the reference data.}
\end{figure}

The resulting CRP shows a clear LOS and some clustering of 
black patches (Fig.~\ref{crp}). The latter occurs due 
to the plateaus in the data.
The next step is to fit a non-parametric function (the 
depth-depth-curve) to the
LOS in the CRP (red curve in Fig.~\ref{crp}). With this
function we are able to adjust the data of the PS~2178-3 core
to the scale of PS~2180-2 (Fig.~\ref{arm_aa}). 

The determination of the depth-depth-function with the
conventional method of visual wiggle matching is based on the
interactive and parallel searching for the same structures in 
the different parameters of both data sets. If the adjustment
does not work in a section of the one parameter, one can 
use another parameter for this section, which allows
the multivariate adjustment of the data sets. The
recognition of the same structures in the data sets requires
a degree of experience. However, human eyes are usually better
in the visual assessment of complex structures 
than a computational algorithm.

\begin{figure}[h]
  \includegraphics[width=8.2cm]{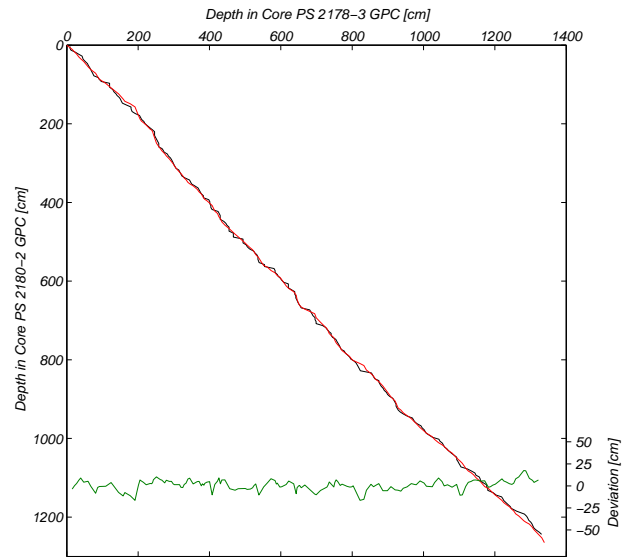}
  \caption[]{\label{depth_depth_curves}
Depth-depth-curves. In black the curve gained with the CRP,
in red the manually matching result. The green curve shows the 
deviation between both results.}
\end{figure}

Our depth-depth-curve differs slightly from the curve which was 
gained by the visual wiggle matching 
(Fig.~\ref{depth_depth_curves}). However, despite our (still)
weak algorithm used to fit the non-parametric adjustment function
to the LOS, we obtained a good result of adjusted data
series. If they are well adjusted, the correlation 
coefficient between the parameters of the adjusted 
data and the reference data should not vary so much.
The correlation coefficients between the reference and
adjusted data series is about $0.70$ -- $0.80$, where the
correlation coefficients of the interactive rescaled 
data varies from $0.71$ -- $0.87$ (Tab.~\ref{tab_cor}). 
The $\chi^2$ measure of the correlation coefficients emphasizes 
more variation for the wiggle matching
than for the CRP rescaling.

\begin{figure}
  \includegraphics[width=8.3cm]{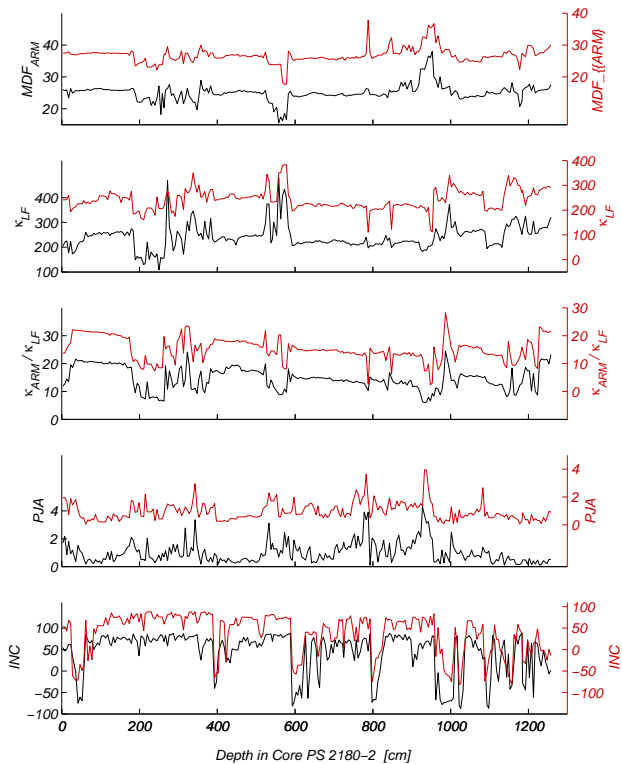}
  \caption[]{\label{daten_aa}
The adjusted marine sediment parameters. The construction
of the CRP was done with the normalized parameters. In this
plots we show the parameters, which are not normalized.}
\end{figure}

\begin{table}[htbp]
\caption{Correlation coefficients $\varrho_{1,\,2}$ between adjusted data
and reference data and their $\chi^2$ deviation. The correlation of 
the interactive adjusted data varies more than the automatic adjusted data. 
The data length is $N=170$ (wiggle matching) and $N=250$
(CRP matching). The difference between the both correlation 
coefficients $\varrho_1$ and $\varrho_2$ is significant at a 99\,\% 
significance level, when
the test measure $\hat z$ is greater than $z_{0.01}=2.576$. }\label{tab_cor}
\begin{center}
\begin{tabular}{lrrr}
\hline
Parameter 	&$\varrho_1$, wiggle matching	&$\varrho_2$, CRP matching&$\hat z$\\
\hline
$ARM$			&0.8667			&0.7846	&6.032\\
$M\!D\!F_{ARM}$		&0.8566			&0.7902	&4.791\\
$\kappa_{L\!F}$		&0.7335			&0.7826	&2.661\\
$\kappa_{ARM}/\kappa_{L\!F}$	&0.8141		&0.8049	&0.614\\
$P\!J\!A$		&0.7142			&0.6995	&0.675\\
$I\!N\!C$		&0.7627			&0.7966	&1.990\\
\hline
$\chi^2$		&141.4			&49.1	&\\
\hline
\end{tabular}
\end{center}
\end{table}

\section{Discussion}

Cross recurrence plots (CRP) reveal similarities in the states
of the two systems. A similar trajectory evolution gives a diagonal
structure in the CRP. An additional time dilatation or compression
of one of these similar trajectories causes a distortion of
this diagonal structure (Fig.\,\ref{crp.variation}). This effect
is used to look into the synchronization between both systems.
Synchronized systems have diagonal structures along and in the
direction of the main diagonal in the CRP. Interruptions of these
structures with gaps are possible because of variations in the
amplitudes of both systems. However, a loss of synchronization
is viewable by the distortion of this structures along the main
diagonal (LOS). By fitting
a non-parametric function to the LOS one allows to re-synchronization
or adjustment to both systems at the same time scale. Although
this method is based on principles from deterministic dynamics,
no assumptions about the underlying systems has to be made
in order for the method to work. 

The first example shows the obvious relationship between the 
LOS and the time domains of the
considered time series. The squared increasing of the 
frequency of the second harmonic function causes a parabolic 
LOS shape in the CRP (Fig.\,\ref{crp.bsp1}). 
Finally, with 
this LOS we are able to rescale the second function to the
scale of the first harmonic function (Fig.\,\ref{rescaled.bsp1}). 
Some differences in
the amplitude of the result are caused by the algorithm used
in order to extract the LOS from the CRP. However, our concern is to 
focus on the distorted main diagonal and their relationship with
the time domains. 

The second example deals with real geological data and
allows a comparison with the result of the conventional method 
of visual wiggle matching. The visual comparison of the
adjusted data shows a good concordance with the reference
and the wiggle matched data (Fig.~\ref{arm_aa} and \ref{daten_aa}).
The depth-depth-function differs up to 20 centimeters from the
depth-depth-function of the wiggle matching.
The correlation coefficients between the CRP adjusted data and
the reference data varies less than the correlation coefficients
of the wiggle matching. However, the correlation coefficients
for the CRP adjusted data are smaller than these for the
wiggle matched data. Although their correlation
is better, it seems that the interactive method
does not produce a balanced adjusting, whereas the automatic matching 
looks for a better balanced adjusting. 

These both examples exhibits the ability to work with smooth and
non-smooth data, whereby the result will be better for smooth
data. Small fluctuations in the non-smooth data can be handled 
by the LOS searching algorithm. Therefore, smoothing strategies, 
like smoothing or parametrical fit of the LOS,
are not necessary. The latter would damp one advantage of this 
method, that the LOS is yielded as a non-parametrical function.
A future task will be the optimization of the LOS searching algorithm,
in order to get a clear LOS even if the data are non-smooth.
Further, the influence of dynamical noise to the result will be studied. 
Probably, this problem may be bypassed by a suitable LOS searching 
algorithm too.

Our method has conspicuous similarities with the 
method of sequence slotting described by \cite{thompson89}.
The first step in this method is the calculation of a distance 
matrix similar to our Eq.~\ref{eqCRP}, which allows the use 
of multivariate data sets. \cite{thompson89} referred to the distance measure
as dissimilarity. It is used to determine the alignment 
function in such a way that the sum of the dissimilarities 
along a path in the distance matrix is minimized. This approach is
based on dynamic programming methods which were mainly developed
for speech pattern recognition in the 70's \citep[e.\,g.][]{sakoe78}.
In contrast, RPs were developed to visualize the phase space
behaviour of dynamical systems. Therefore, a threshold was introduced 
to make recurrent states visible. The involving of a fixed amount
of nearest neighbours in the phase space and the possibility to 
increase the embedding dimensions distinguish this approach from
the sequence slotting method.

\section{Conclusion}
The cross recurrence plot (CRP) can contain information about the synchronization
of data series. This is revealed by the distorted main diagonal, which
is called {\it line of synchronization (LOS)}. After isolating this LOS
from the CRP, one obtains a non-parametric rescaling function. With this
function, one can synchronize the time series. The underlying
more-dimensional phase space allows to include more than one parameter
in this synchronization method, as it usually appears in geological
applications, e.\,g.~core synchronization. The comparison of
CRP adjusted geophysical core data with the conventionally 
visual matching shows an acceptable reliability level of the
new method, which can be further improved by a better method for 
estimating the LOS. The advantage is the automatic,
objective and multivariate adjustment.
Finally, this method of CRPs can open a wide range of 
applications as scale adjustment, phase synchronization and pattern recognition
for instance in geology, molecular biology and ecology.

\begin{acknowledgements}
The authors thank Prof.~J\"urgen Kurths and Dr.~Udo Schwarz
for continuing support and discussion. This work was supported
by the special research programme 1097 of the German Science Foundation 
(DFG).
\end{acknowledgements}

\section*{Appendix: An Algorithm to Fit the LOS}

In order to implement a recognition of the LOS
we have used the following simple two-step algorithm. 
Denote all recurrence points by
$r_{i_{\tilde \alpha}, j_{\tilde \beta}}$
($\tilde \alpha, \tilde \beta=1, 2, \ldots$) and the recurrence points
lying on the LOS by $r_{i_{\alpha},j_{\beta}}$
($\alpha, \beta=1, 2, \ldots$). Before the
common algorithm starts, find the recurrence point $r_{i_1,j_1}$ 
next to the axes origin. In the first step, the next 
recurrence point $r_{i_{\tilde \alpha},j_{\tilde \beta}}$ after a previous 
determined recurrence point $r_{i_{\alpha},j_{\beta}}$ is to be determined. 
This is carried out by
a step-wise increasing of a squared $(w \times w)$ sub-matrix, 
wherein the previous recurrence point is at the $(1,1)$-location. The size
$w$ of this sub-matrix increases step-wise until it meets a
new recurrence point or the margin of the CRP.
When a next recurrence point 
$r_{i_{\tilde \alpha},j_{\tilde \beta}}=r_{i_{\alpha}+\delta i,j_{\beta}+\delta j}$ 
($\delta i=w$ or $\delta j=w$) in the $x$-direction ($y$-direction)
is found, the second step looks if there are following 
recurrence points in $y$-direction 
($x$-direction). If this is true (e.\,g.~there are a cluster 
of recurrence points)  increase further the sub-matrix 
in $y$-direction ($x$-direction) until a predefined size 
$(w+d\tilde x) \times (w+d\tilde y)$
($d\tilde x<dx, d\tilde y<dy$) or until no new recurrence points are
met. This further increasing of the sub-matrix is done for the both
$x$- and $y$-direction. Using $d\tilde x, d\tilde y$ we compute 
the next recurrence point $r_{i_{\alpha+1},j_{\beta+1}}$ by 
determination of the center of mass of the cluster of recurrence points
with $i_{\alpha+1}=i_{\alpha}+(d\tilde x + \delta i)/2$ and 
$j_{\beta+1}=j_{\beta}+(d\tilde y + \delta j)/2$.
The latter avoids the fact that the algorithm is driven around 
widespread areas of recurrence points. Instead of this, the
algorithm locates the LOS within these areas. (However, the introducing
of two additional parameter $dx$ and $dy$ is a disadvantage which
should be avoided in further improvements of this algorithm.)
The next step is to set the recurrence point 
$r_{i_{\alpha+1},j_{\beta+1}}$ to a new 
start point and to begin with the step one 
in order to find the next recurrence point. 
These steps are repeated until the end of the RP is reached.

We know that this algorithm is merely one of many
possibilities. The following criteria should be met in order
to obtain a good LOS. The amount of targeted recurrence points by
the LOS $N_1$ should converge to the maximum and the amount of gaps in
the LOS $N_0$ should converge to the minimum. An analysis with various
estimated LOS confirms this requirement. The correlation between 
two LOS-synchronized data series arises with $N_1$
and with $1/N_0$ (the correlation coefficient correlates 
most strongly with the ratio $N_1/N_0$).

The algorithm for computation of the CRP and recognition of the
LOS are available as Matlab programmes on the WWW:
http://www.agnld.uni-potsdam.de/\texttildelow marwan.

\bibliographystyle{egs}
\bibliography{../mybibs}

\end{document}